\documentclass[twocolumn,twoside,slac_two]{revtex4}
\usepackage{graphicx}
\usepackage{fancyhdr}
\begin{document}

\title{Environment and properties of emitting electrons in blazar jets: Mrk\,421 as a laboratory}

%

\author{Nijil Mankuzhiyil}
\affiliation{INFN Trieste and  Universit\`{a} di Udine, via delle Scienze 208, I-33100 Udine, ITALY}
\author{Stefano Ansoldi}
\affiliation{International Center for Relativistic Astrophysics (ICRA), Rome and  Universit\`{a} di Udine, via delle Scienze 208, I-33100 Udine, ITALY }

\author{Massimo Persic}
\affiliation{INAF-Trieste, via G.\,B.\,Tiepolo 11, I-34143 Trieste, ITALY}

\author{Fabrizio Tavecchio}
\affiliation{INAF-Brera, via E.\,Bianchi 46, I-23807 Merate, ITALY}

\begin{abstract}

Here we report our recent study on the spectral energy distribution (SED) of the high frequency BL\,Lac object Mrk\,421 in different luminosity states. We used a full-fledged $\chi^2$-minimization procedure instead of more commonly used "eyeball" fit to model the observed flux of the source (from optical to very high energy), with a Synchrotron-Self-Compton (SSC) emission mechanism. Our study shows that the synchrotron power and peak frequency remain constant with varying source activity, and the magnetic field ($B$) decreases with the source activity while the break energy of electron spectrum ($\gamma_{br}$) and the Doppler factor ($\delta$) increase. Since a lower magnetic field and higher density of electrons result in increased electron-photon scattering efficiency, the Compton power increases, so does the total emission.

\end{abstract}

\maketitle

\thispagestyle{fancy}


\section{Introduction}

Active galactic nuclei (AGN) involve the most powerful, steady sources of luminosity in the Universe. It is believed that the center core of AGN consist of super massive black hole (SMBH) surrounded by an accretion disk. In some cases powerful collimated jets are found in AGN, perpendicular to the plane of accretion disk. The origin of jets are still unclear. AGNs whose jets are viewed at a small angle to its axis are called blazars. 

The overall (radio to $\gamma$-ray) spectral energy distribution (SED) of blazars shows 
two broad non-thermal continuum peaks. 
The low-energy peak is thought to arise from electron synchrotron emission. The leptonic 
model suggests that the second peak forms due to inverse Compton emission. This can be due 
to upscattering, by the same non-thermal population of electrons responsible for the 
synchrotron radiation, and synchrotron photons (Synchrotron Self Compton: SSC)~\cite{Mar+92}. 

Blazars often show violent flux variability,  that may or may not appear correlated in the 
different energy bands.  Simultaneous observation are then crucial to understand the physics behind variability.

\section{$\chi^2$-minimized SED fitting}

In this section we discuss the code that we have used to obtain an estimation
of the characteristic parameters of the SSC model. The SSC model assumes a spectrum
for the accelerated electron density $k$, which is a broken power law with exponents
$n _{1}$ and $n _{2}$. The minimum, maximum and break Lorentz factors for the electrons
are usually called $\gamma _{\mathrm{min.}}$, $\gamma _{\mathrm{max.}}$ and $\gamma _{\mathrm{break}}$
respectively. The emitting region is considered to be a blob of radius $R$ moving with Doppler factor
$\delta$ with respect to the observer in a magnetic field of intensity $B$.
The model is thus characterized by nine free parameters.

\begin{table}
    \hrule
    \begin{eqnarray}
        && \mbox{DEF: SSC parameters initial values set-up} \nonumber\\
        && \mbox{calculate initial } \chi ^{2} \mbox{ value, change parameters} \nonumber\\
        && \mathrm{LOOP:} \nonumber\\
        &&      \qquad \mbox{calculate } \chi ^{2} \mbox{ for modified parameters }\nonumber\\
        &&      \qquad \mbox{if } \chi ^{2} \mbox{ has increased:} \nonumber\\
        &&      \qquad \qquad \mbox{\it\small{}we are moving away from a minimum } \nonumber\\
        &&      \qquad \qquad \qquad \Rightarrow \mbox{\footnotesize{}change parameters, increase weight} \nonumber\\
        &&      \qquad \qquad \qquad \phantom{\Rightarrow} \mbox{\footnotesize{}of steepest descent method and reset} \nonumber\\
        &&      \qquad \qquad \qquad \phantom{\Rightarrow} \mbox{\footnotesize{}negligible decrease amount counter} \nonumber\\
        &&      \qquad \mbox{if } \chi ^{2} \mbox{ has decreased:} \nonumber\\
        &&      \qquad \qquad \mbox{\it\small{}we are moving toward a minimum } \nonumber\\
        &&      \qquad \qquad \qquad \Rightarrow \mbox{\footnotesize{}change parameters and increase} \nonumber\\
        &&      \qquad \qquad \qquad \phantom{\Rightarrow} \mbox{\footnotesize{}weight of inverse Hessian method} \nonumber\\
        && \mbox{UNTIL: } \chi ^{2} \mbox{ decreases by a negligible amount} \nonumber \\
        && \mbox{\phantom{UNTIL: }} \mbox{for the fourth time} \nonumber
    \end{eqnarray}
    \hrule
\caption{\label{tab:LevMarmet} The $\chi^2$ minimization algorithm.}
\end{table}

In the present work we have kept
$\gamma _{\mathrm{min.}}$ fixed and equal to unit, which is a satisfactory
approximation already used in the literature. The determination of the remaining eight parameters has been performed
by finding their best values and uncertainties from a $\chi ^{2}$ minimization
in which multi-frequency experimental points have been fitted to the SSC spectrum
modelled as in \cite{Tav+98}. Minimization has been performed using the
Levenberg-Marquardt method \cite{Press+94}, which is an efficient standard for
non-linear least-squares minimization that smoothly interpolates between two different
minimization approaches, namely the inverse Hessian method and the steepest
descent method. For completeness, we briefly present the pseudo-code for
the algorithm in table I.

A crucial point in our implementation is that from \cite{Tav+98}
we can only obtain a numerical approximation to the SSC spectrum, in the form of a
sampled SED. On the other hand, from table I, we understand that at each step
the calculation of the $\chi^{2}$ requires the evaluation of the SED for all the
observed frequencies. Although an observed point will likely not be one of the sampled points
coming from \cite{Tav+98}, it will fall between two sampled points, so that interpolation
can be used to approximate the value of the SED\footnote{The sampling of the
SED function coming from \cite{Tav+98} is dense enough, so that, with respect to other
uncertainties, the one coming from this interpolation is negligible.}.

At the same time, the Levenberg-Marquardt method requires the calculation of the partial derivatives
of $\chi^{2}$ with respect to the SSC parameters. These derivatives have also been obtained
numerically by evaluating the incremental ratio of the $\chi^{2}$ with respect to a sufficiently
small, dynamically adjusted increment of each parameter. This method could have introduced a potential
inefficiency in the computation, due to the recurrent need to evaluate the SED at many, slightly different points
in parameter space, this being the most demanding operation in terms of CPU time. For this reason
we set up the algorithm to minimize the number of calls to \cite{Tav+98} across
different iterations. The $\chi^2$ fit during different iterations are shown in Fig.\,1.

\begin{table}[t!] 
\begin{center}
\caption{Data sets used in this study. The observation period of each state can be found at Fig.\,2.}
\begin{tabular}{l c c c}
\hline \textbf{State} & \textbf{Instruments} & \textbf{References}
\textbf{}

\\
\hline 1. & XMM-$Newton$ & \cite{flaring_state126} \\
& Whipple, MAGIC & \\

 2. & XMM-$Newton$ & \cite{flaring_state126} \\
& Whipple, MAGIC & \\

 3. & KVA, WIYN, RXTE & \cite{{flaring_state3}} \\
& Whipple, HEGRA-CT\,1 & \\

 4. & Boltwood, RXTE & \cite{flaring_state49} \\
& Whipple & \\

 5. & Havard-Smithsonian & \cite{Foss+08} \\
& RXTE, Whipple & \\

 6. & XMM-$Newton$ & \cite{flaring_state126} \\
& VERITAS & \\

 7. & Havard-Smithsonian & \cite{Foss+08} \\
& RXTE, Whipple & \\

 8. & WEBT, $Swift$ & \cite{flaring_state8} \\
& RXTE, VERITAS & \\

 9. & Boltwood, RXTE & \cite{flaring_state49} \\
& Whipple & \\

\hline
\end{tabular}
\label{l2ea4-t1}
\end{center}
\end{table}

\begin{figure*}
\centering
\includegraphics[width=105mm]{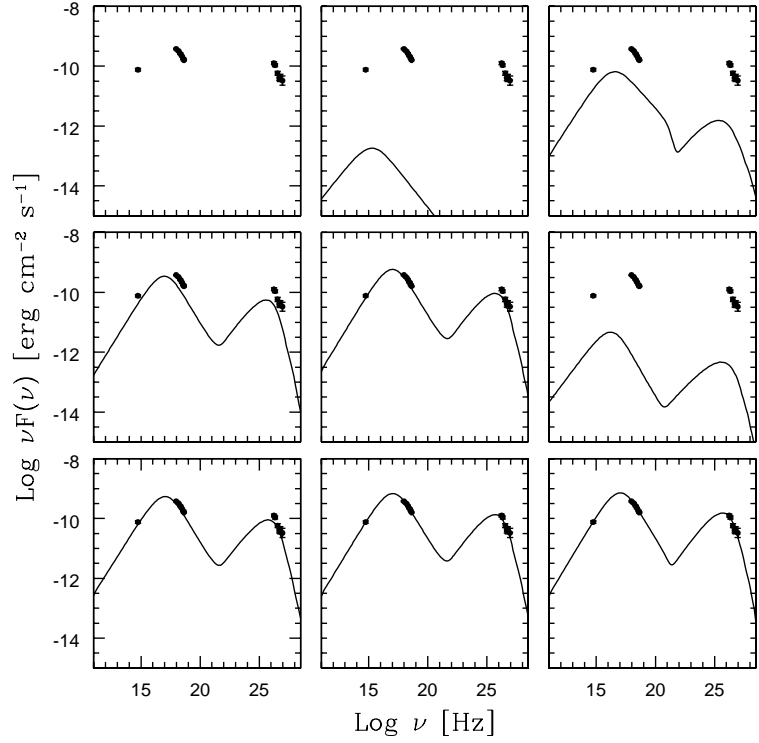}
\caption{Steps during $\chi^2 minimization$ iterations.} \label{JACpic2-f1}
\end{figure*}

\section{Application and results}

In order to study the behavior of parameters with source activity, we choose Mrk\,421 (table II), considering the larger availability of MWL data sets and the lower redshift, hence less uncertainty after EBL correction of VHE data.
The $\chi^2$ fitted SEDs are shown in Fig.\,2.

In addition to the $\chi^2$ test, we also checked the goodness of the fit using the Kolmogorov-Smirnov (KS) test. Considering the occurrence of different physical processes (synchrotron and inverse Compton, at substantially different energies), and the different quality of low- and high-energy data, we used a \emph{piecewise KS test}, \emph{i.e.} we applied the KS test separately to low- and high-energy data. Then the KS test always confirms  that the fit residuals are normal at 5\% confidence level.

Our results suggest that in Mkn\,421, $B$ decreases with source activity whereas $\gamma_{\rm break}$ and $\delta$ increase (Fig.\,3 top). This can be interpreted in a frame where the synchrotron power and peak frequency remain constant with varying source activity by decreasing magnetic field and increasing the number of low energy electrons. This mechanism results in an increased electron-photon scattering efficiency and hence in an increased Compton power. Other emission parameters appear uncorrelated with source activity. In Fig.\,3 (bottom), the $B$-$\gamma_{\rm break}$ anti-correlation results from a roughly constant synchrotron peak frequency. The $B$-$\delta$ correlation suggests that the Compton emission of Mkn\,421 is always in the Thomson limit. The $\delta$-$\gamma_{\rm break}$ correlation is an effect of the constant synchrotron and Compton frequencies of the radiation emitted by a plasma in bulk relativistic motion towards the observer. 

More detailed description of this work is available in \cite{Mankuzhiyil2011}.

\begin{figure*}
\centering
\includegraphics[width=105mm]{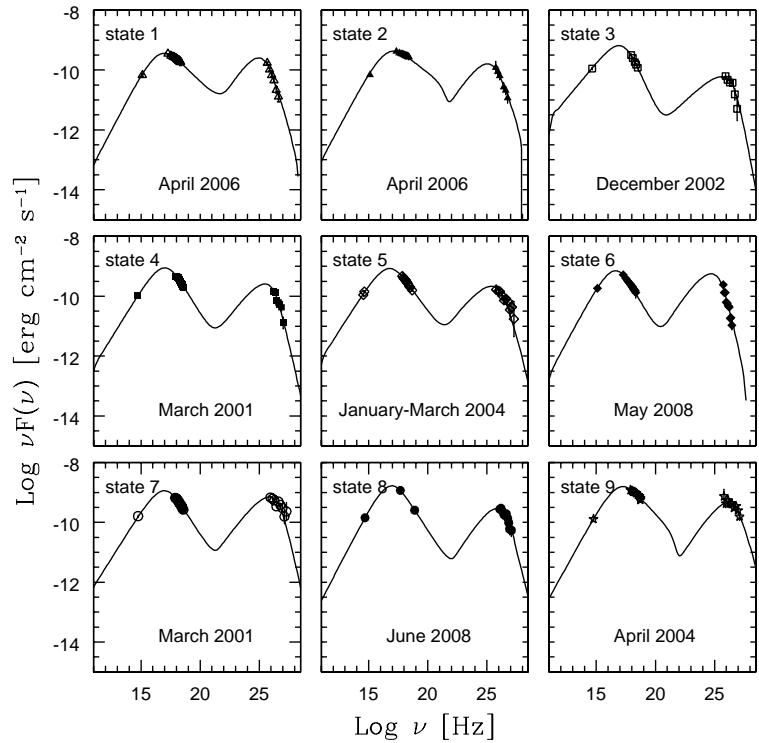}
\caption{Mrk\,421 MWL data sets and corresponding $\chi^2$ minimized SED fits.} \label{JACpic2-f1}
\end{figure*}

\begin{figure*}
\centering
\includegraphics[width=135mm]{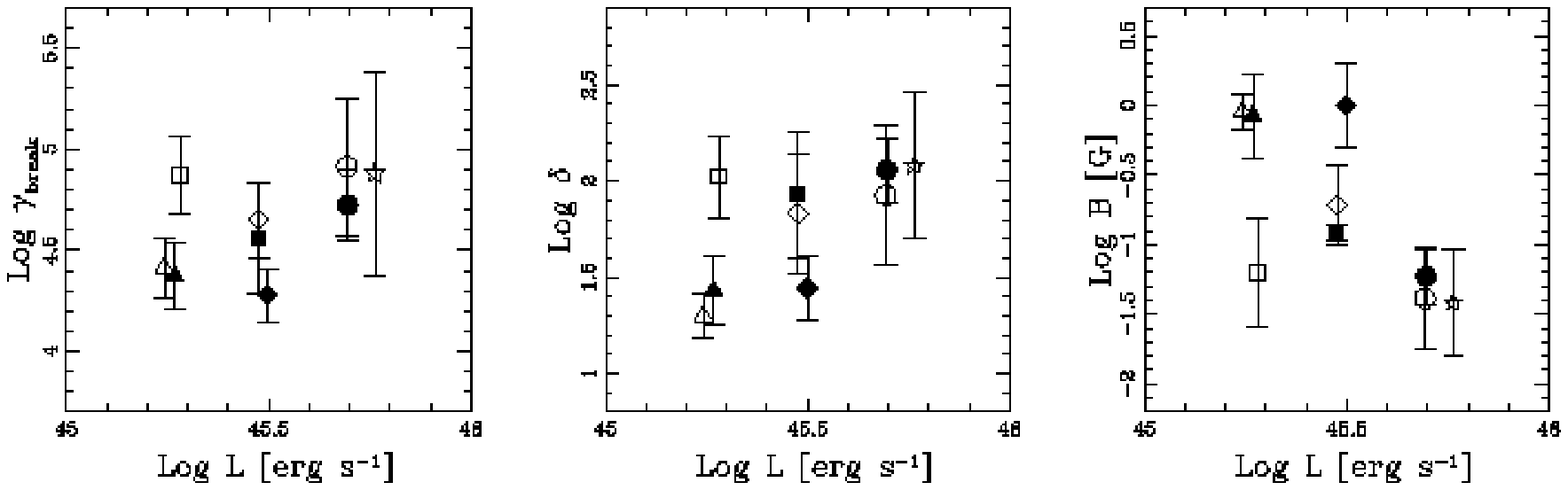}
\includegraphics[width=135mm]{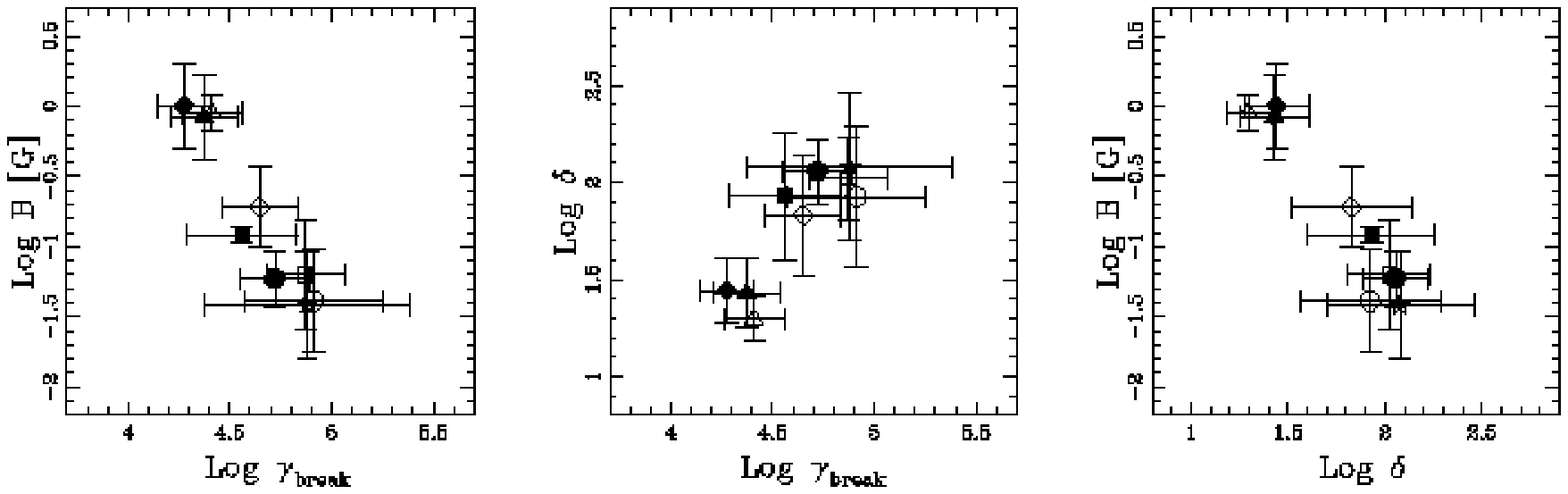}
\caption{{\it Top.} Variations of the fitted parameters - $B$, $\delta$, and $\gamma_{\rm br}$ - as a function of luminosity. The other SSC parameters show a scatter plot with the luminosity. Parameters and its uncertainty can be found at \cite{Mankuzhiyil2011}.
{\it Bottom.} Correlations between $B$, $\delta$, and $\gamma_{\rm br}$.} \label{JACpic2-f1}
\end{figure*}

\bigskip 

\end{document}